\documentclass{elsart}

\usepackage{graphicx}
\usepackage{natbib}

\newcommand{\lsim}{\,\lower2truept\hbox{${<\atop\hbox{\raise4truept\hbox{$\sim$}}}$}\,}
\newcommand{\gsim}{\,\lower2truept\hbox{${>\atop\hbox{\raise4truept\hbox{$\sim$}}}$}\,}

\begin{document}

\begin{frontmatter}

\title{Weighting CMB and Galactic synchrotron polarisation}

\author{Carlo Baccigalupi}

\address{SISSA/ISAS, Via Beirut 4, 34014 Trieste, Italy}

\begin{abstract}
We review the present knowledge of the diffuse Galactic 
synchrotron emission in polarisation. 
At microwave frequencies, we assess the expected contamination 
to the CMB polarisation angular power spectrum, for $E$ and $B$
modes, as expected after the WMAP first year measurements. 
\end{abstract}


\end{frontmatter}

\section{Introduction}
\label{intro}

An extraordinary improvement in Cosmic Microwave Background (CMB) 
observations is presently ongoing (see Bennett \etal 2003 and references 
therein). Several balloon-borne and ground-based observations 
map CMB anisotropies on angular scales going 
from a few arcminutes to tens of degrees; the Wilkinson Microwave 
Anisotropy Probe\footnote{\tt map.gsfc.nasa.gov} (WMAP) satellite 
released recently the first year full sky maps, with angular resolution 
$\gsim 14$ and a sensitivity of the order of ten $\mu$K, on a frequency 
range extending from 22 to 90 GHz. The {\sc Planck} 
satellite\footnote{\tt astro/estec.esa.nl/SA-general/Projects/Planck} 
will provide total intensity and polarisation full sky maps with 
resolution $\gsim 5'$ and a sensitivity of a few $\mu$K, on nine channels 
in the frequency range 30-857 GHz. 

Measurements of CMB polarisation are still at the beginning. As it 
is well known (see Zaldarriaga \& Seljak 1997 and Kamionkowski, Kosowsky \& 
Stebbins 1997), CMB polarisation is conveniently expressed in terms of $E$ 
and $B$ modes, non-local combination of the familiar $Q$ and $U$ Stokes 
parameters. Total intensity $T$ and $E$ components are excited by 
all kinds of cosmological perturbations, namely scalars, vectors and 
tensors, and strongly correlated; the $B$ modes select vectors and 
tensors only. This rich phenomenology, and the hope to reveal tensors, i.e. 
cosmological gravitational waves, is the reason for the great effort 
toward CMB polarisation measurements, despite of the weakness of the 
signal, expected one order of magnitude less than total intensity. 
Recently, a first detection was carried out (Kovac \etal 2002); 
WMAP was also successful in detecting the $TE$ correlation. 

Any CMB observation must control the foreground emission. Polarised 
foregrounds are less known than in total intensity, see De Zotti (2002) 
for reviews. At low CMB frequencies, say 100 GHz or less, the main 
foreground is synchrotron (Haslam \etal 1982, Duncan \etal 1997, 1999, 
Uyaniker \etal 1999), which is the most known in polarisation, and 
the subject of this work. The free-free emission is relevant on 
the same frequencies, and is expected to be negligibly polarised. 
On higher frequencies, the Galactic thermal dust emission dominates, 
and is very poorly known in polarisation (Lazarian \& Prunet 2002). 
Moreover, several populations of extra-Galactic sources are expected at 
all frequencies (see De Zotti 2002), 

In this work we focus on the contamination coming from the 
diffuse polarised Galactic synchrotron emission. In Section 
\ref{guess} we review the current knowledge and forecast 
about this signal. In Section \ref{contamination} we assess 
the level of contamination to the CMB radiation, on $E$ and 
$B$ modes, as expected after the first year WMAP measurements. 

\section{Guessing all sky polarised synchrotron}
\label{guess}

Current radio band observation cover about half of the sky at the 
degree resolution (Brouw \& Spoelstra 1976), as well as low and medium 
Galactic latitudes with $10$ arcminutes resolution 
(Duncan et al. 1997, Uyaniker et al. 1999, Duncan et al. 1999), 
up to $b\simeq 20^{\circ}$. These data allow an estimate of the 
polarised synchrotron fluctuations, and 
in particular their angular power spectrum, up to multipoles 
$\ell\simeq 1500$ (Tucci et al. 2000, Baccigalupi et al. 2001, 
Giardino et al. 2002). 

Giardino et al. (2002) assumed a theoretical polarisation 
degree of $75\%$ correlated with the Haslam et al. (1982) 
template at 408 MHz. The polarisation angle is assumed 
random and obeying the power spectrum derived from 
the high resolution radio observations mentioned above. 
The $Q$ and $U$ simulated templates, reaching a resolution of 
about ten arcminutes, were then scaled in frequencies by considering 
either constant or a space-varying spectral index inferred by 
multi-frequency radio observations. 

On sub-degree angular scales, by analysing the high resolution 
data in the radio band, Tucci \etal (2000) and 
Baccigalupi et al. (2001) found a quite flat slope for the 
polarisation angular power spectrum, 
$C_{\ell}\propto\ell^{-1.5\div 2}$, in agreement with Giardino 
et al. (2002) but with smaller amplitude. 
Baccigalupi et al. (2001) also estimated the power on super-degree 
angular scales, corresponding to multipoles $\ell < 200$, using 
the Brouw \& Spoelstra (1976) data, finding a steeper behavior, 
$C_{\ell}\propto\ell^{-3}$. 

\begin{figure}
\centering
\includegraphics[angle=90,width=6.5cm,height=4cm]{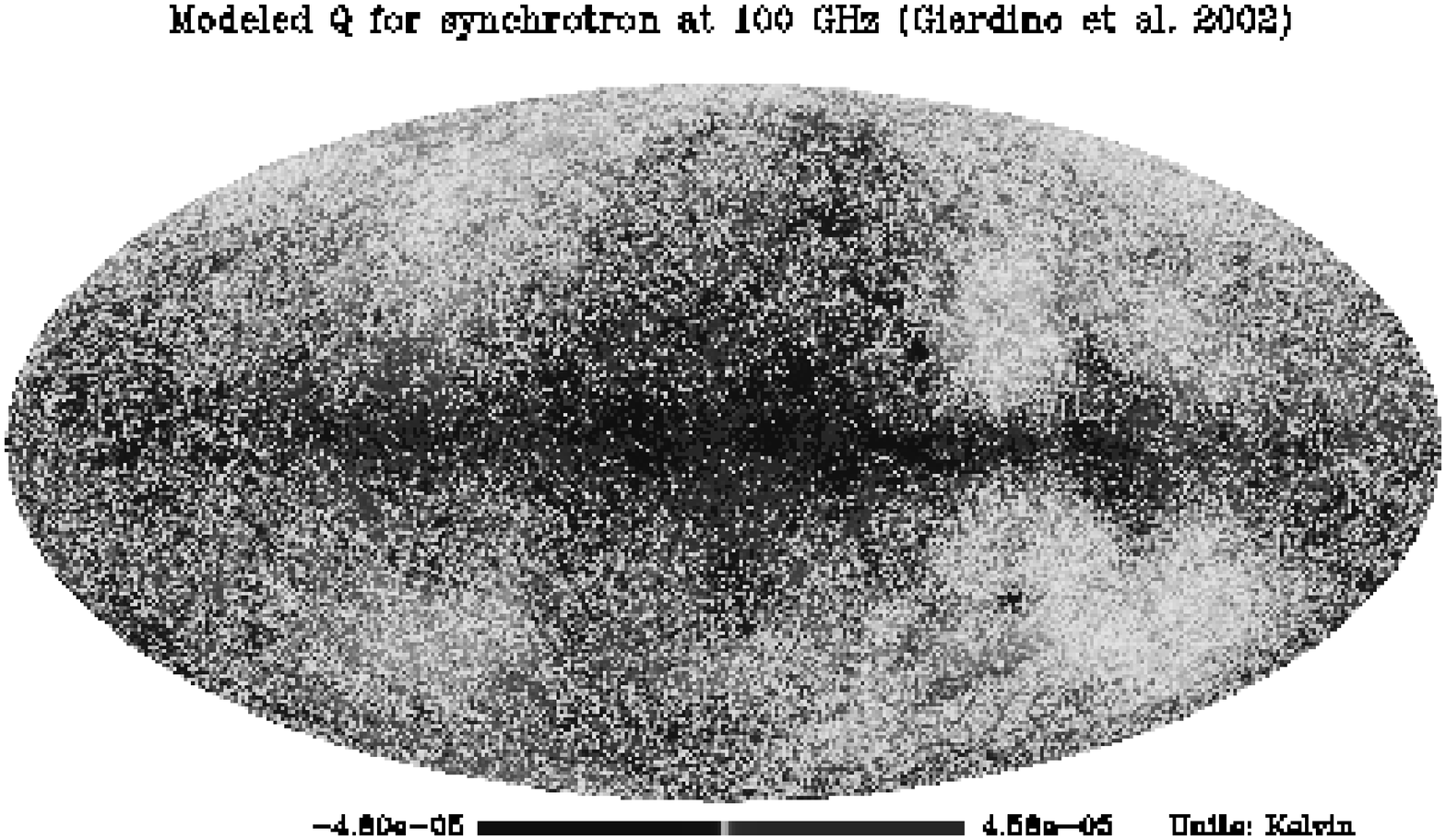}
\includegraphics[angle=90,width=6.5cm,height=4cm]{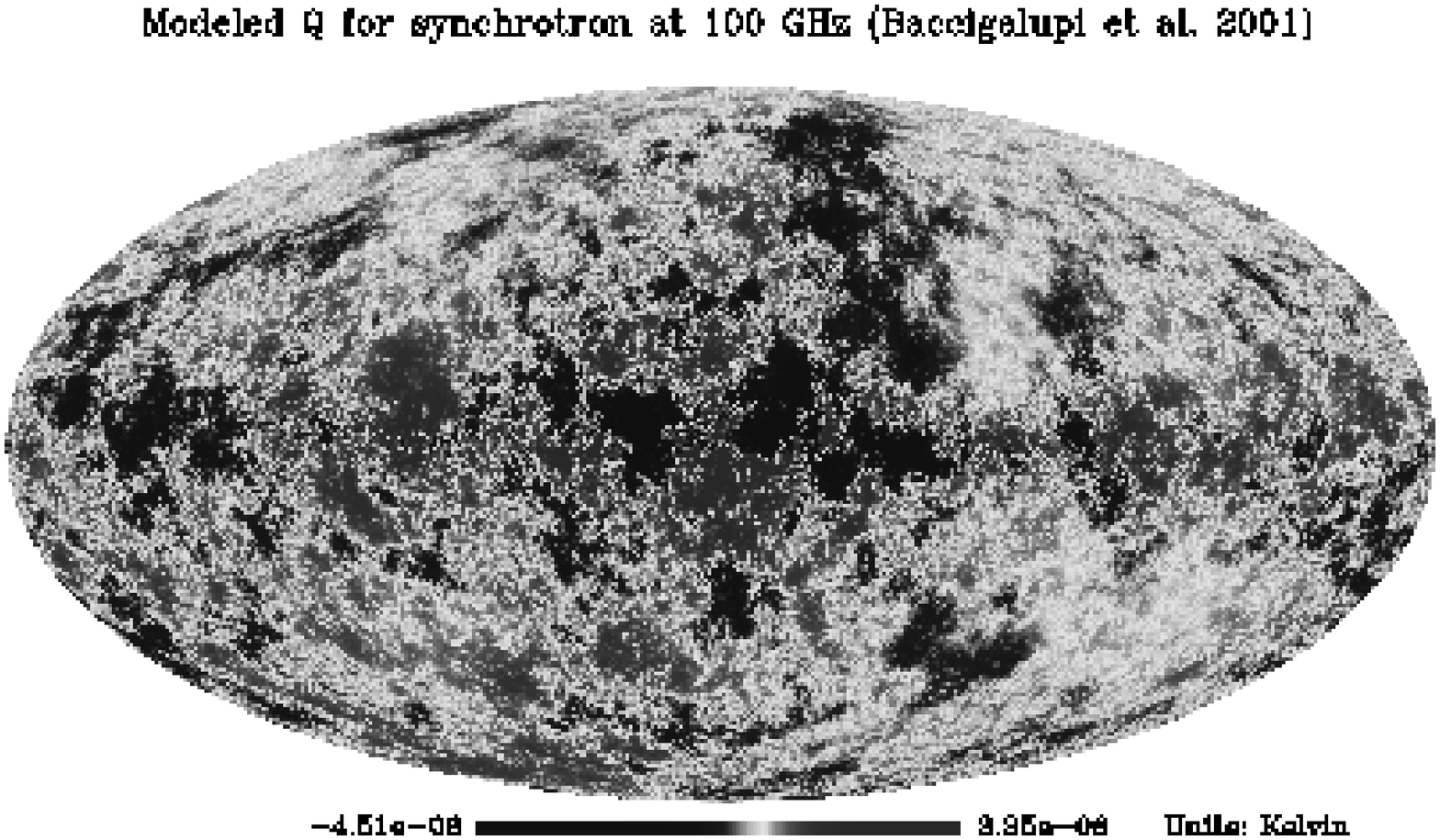}
\caption{Q Stokes parameter for the simulation of Galactic 
synchrotron according to Giardino et al. (2002, left panel) 
and Baccigalupi et al. (2001, right panel).}
\label{synsky}
\end{figure}

The two simulated templates, by Giardino et al. 
(2002), hereafter $S_{G}$, and Baccigalupi et al. (2001), 
hereafter $S_{B}$ (obtained by properly rescaling the power 
of the $S_{G}$ template), are shown in figure \ref{synsky}, 
in a non-linear scale to highlight the behavior at 
high Galactic latitudes; the templates shown are for the $Q$ 
Stokes parameter, the case of $U$ being qualitatively 
analogue, and have been extrapolated at 100 GHz. 
The two templates have roughly the same amplitude, but for $S_{B}$ 
more power on large angular scale can be clearly seen, and we return 
on this in the next Section. 
For completeness, we also show the template for the synchrotron 
spectral index as inferred by Giardino et al. (2002), in figure 
\ref{synspecind}. As it can be seen, the spectral index is far 
from being uniform in the sky, having variations reaching $15\%$ 
on all Galactic latitudes, mostly on large angular scales. 

\begin{figure}
\centering
\includegraphics[width=6.5cm,height=6cm]{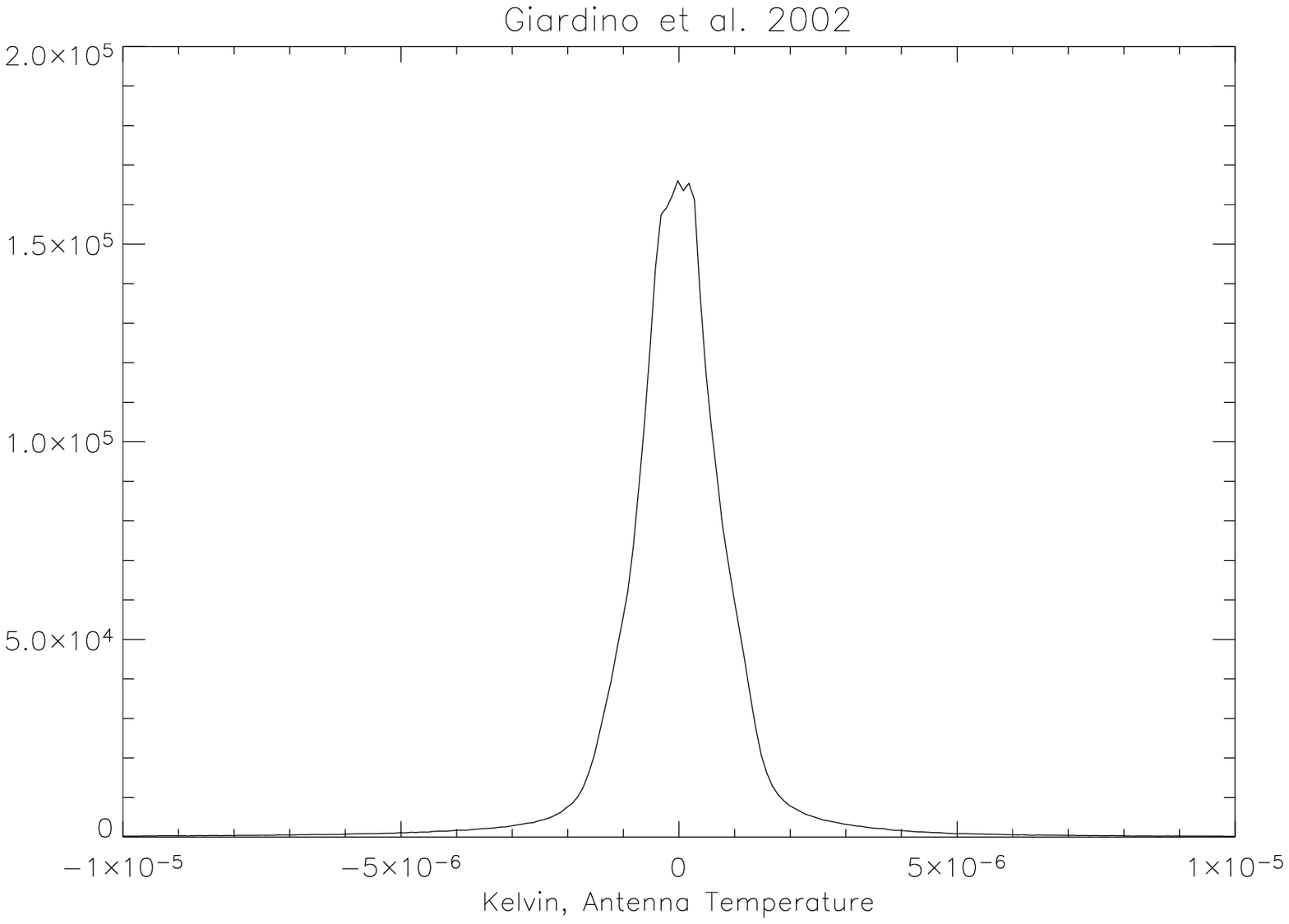}
\includegraphics[width=6.5cm,height=6cm]{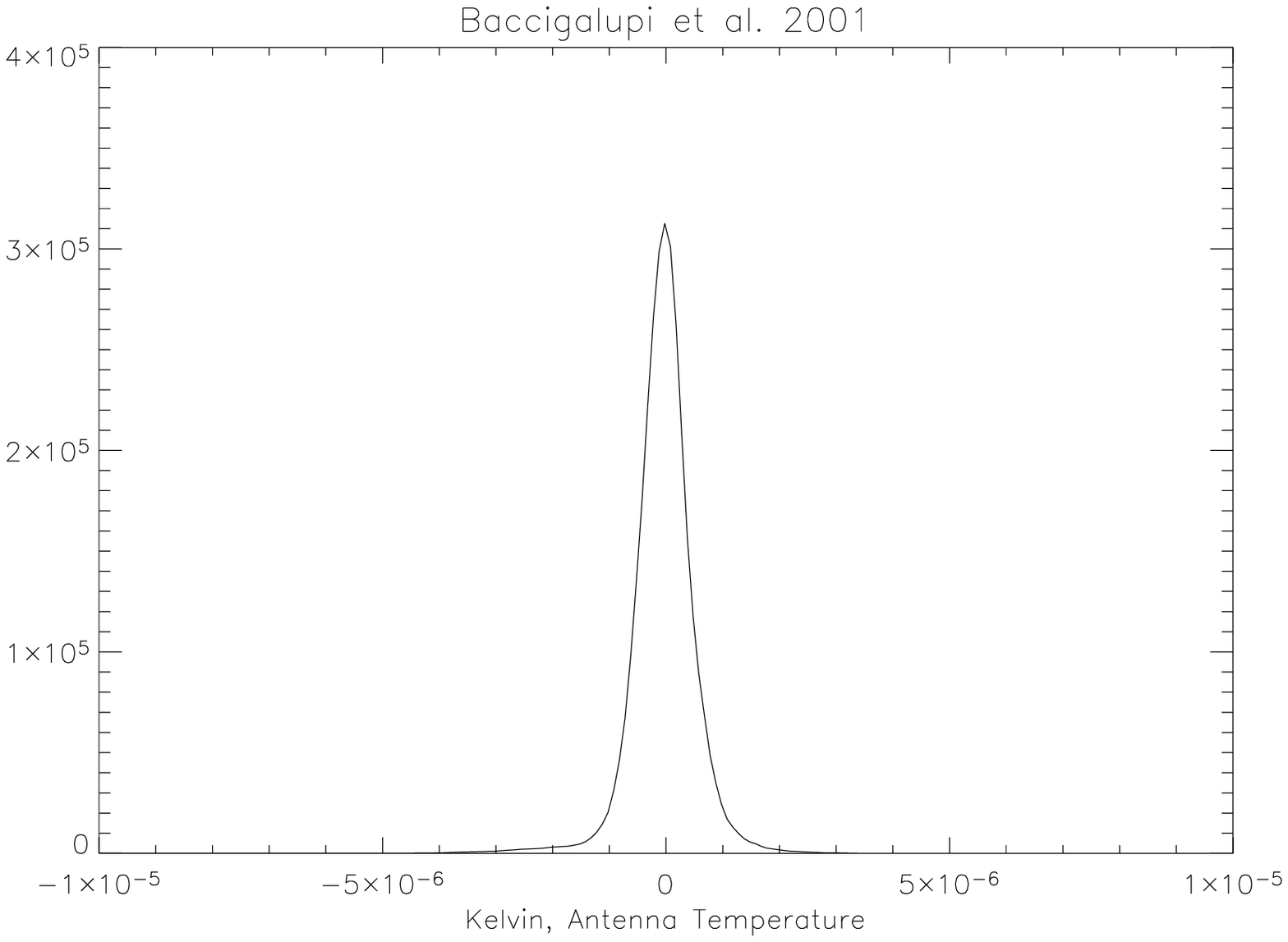}
\caption{Signal distribution of the forecasted diffuse Galactic synchrotron.}
\label{histo}
\end{figure}

Before going to consider the angular power spectrum and 
its relative strength compared to the CMB, it is interesting 
to look at the sky distribution of the forecasted synchrotron 
signal, to be compared with the Gaussianity of the CMB. The signal 
distributions are shown in figure \ref{histo}; as expected for a 
Galactic signal, they are far from Gaussianity, and exhibit a marked 
super-Gaussian behavior. Note that this feature could be useful to 
reduce the contamination to the CMB, since recently proposed 
algorithms rely on the statistical independence of the signals 
to recover (Maino \etal 2002). 

\section{Forecasted CMB contamination}
\label{contamination}

As a CMB template, we generate a realization of the expected 
polarisation signal corresponding to the WMAP measurements, more precisely 
a flat Friedmann Robertson Walker (FRW) model according to table 7 in 
Spergel et al. (2003): 
\begin{equation}
\label{CMB}
h=0.72\ ,\ \Omega_{\Lambda}=0.7\ ,\ \Omega_{b}h^{2}=0.0228
\ ,\ \tau =0.117\ ,\ n_{S}=0.96\ .
\end{equation}
Moreover, we allow for a consistent background of gravitational 
waves, assumed to have a power which is $30\%$ with respect to 
the scalar one, with spectral index given by $-0.044$, according to 
the simplest inflationary prescription. 

\begin{figure}
\centering
\includegraphics[angle=90,width=7cm,height=4.5cm]{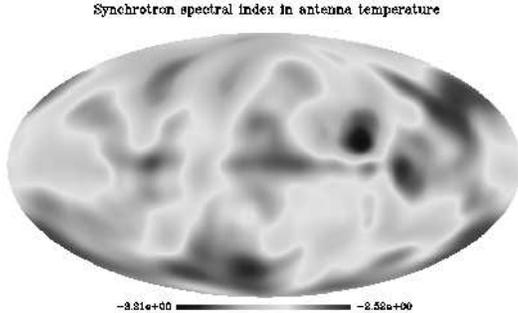}
\caption{Map of synchrotron spectral index as reported in Giardino 
et al. (2002).}
\label{synspecind}
\end{figure}

\begin{figure}
\centering
\includegraphics[width=13cm,height=6.5cm]{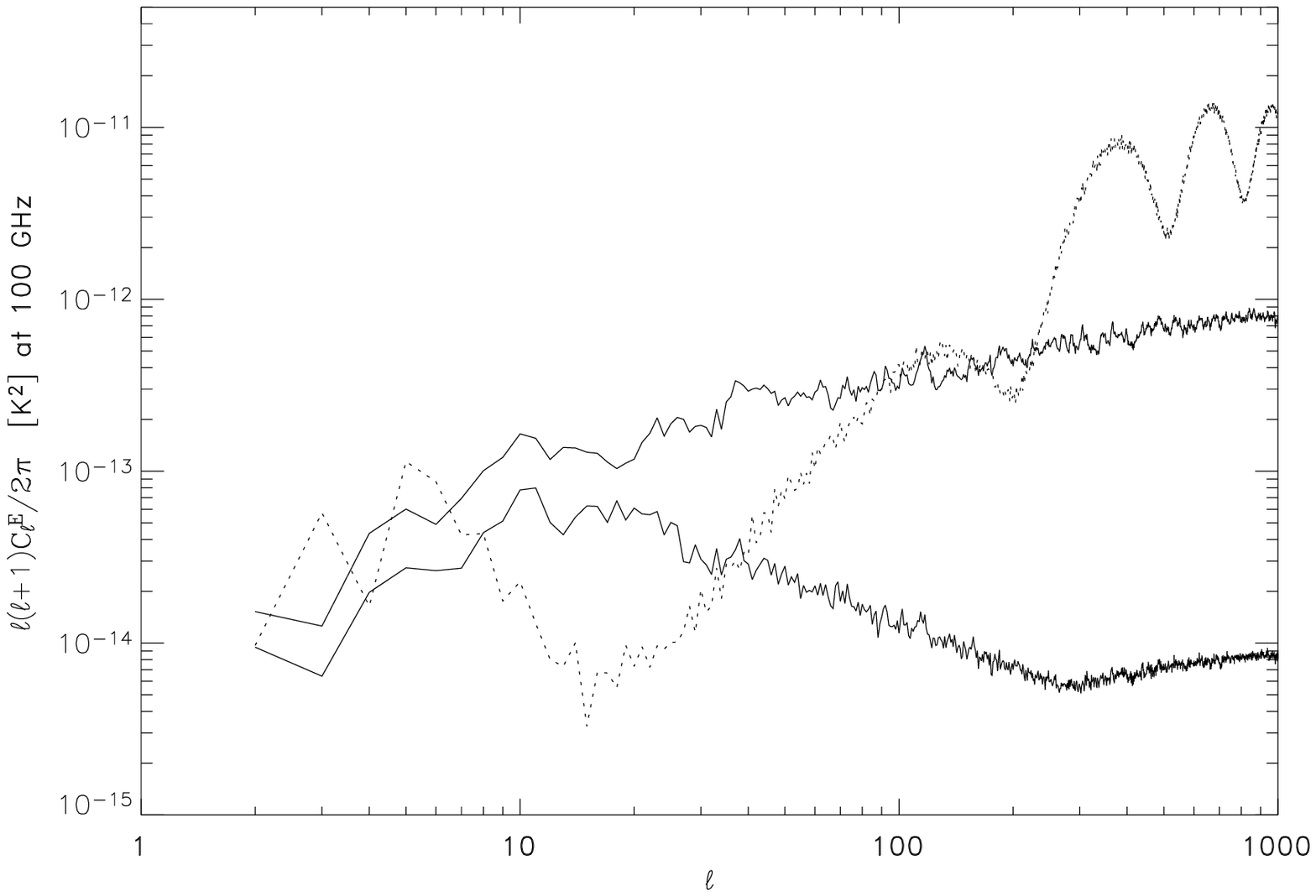}
\includegraphics[width=13cm,height=6.5cm]{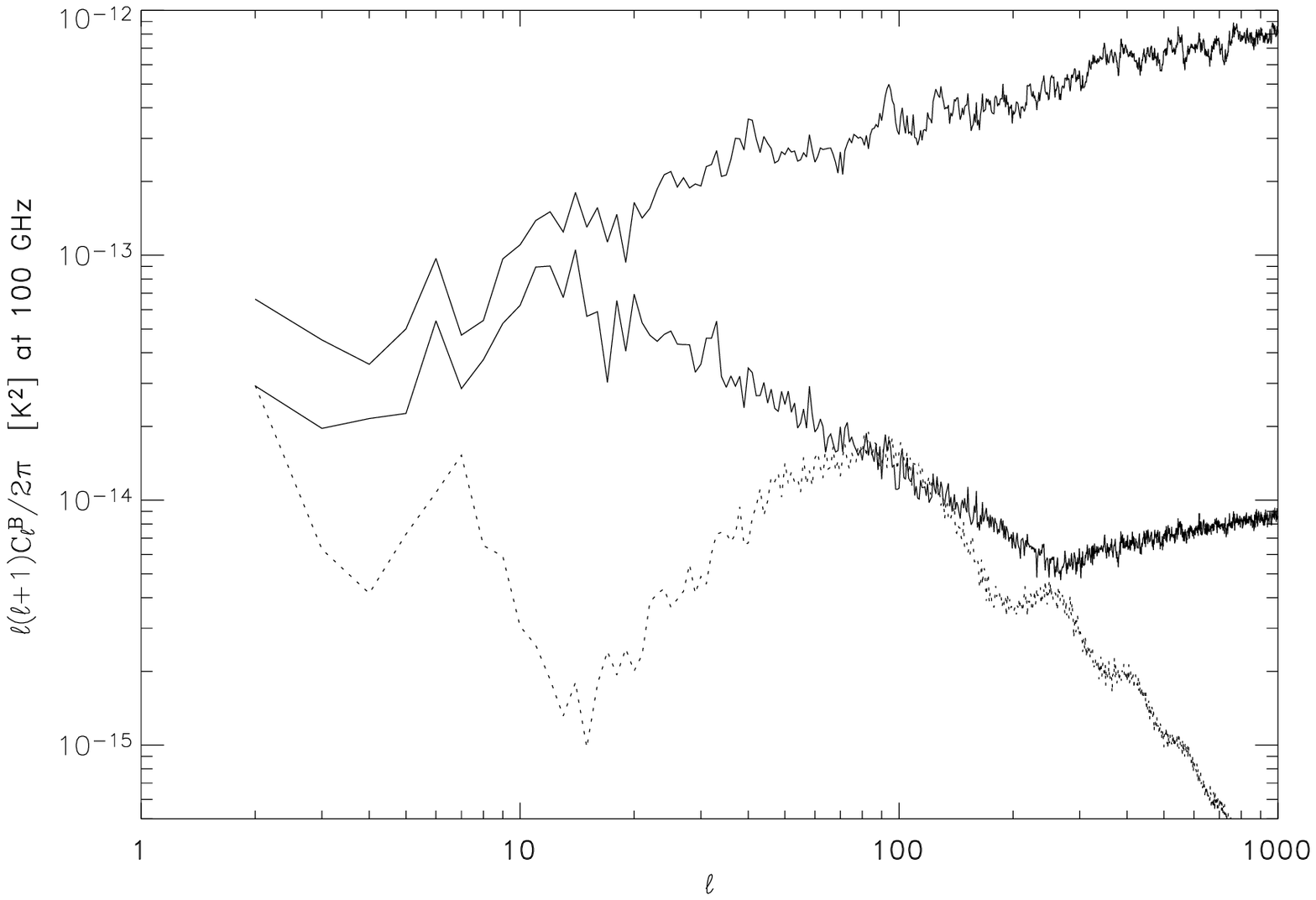}
\caption{All sky CMB $E$ and $B$ modes vs. synchrotron at 100 GHz.}
\label{clCMBsyn}
\end{figure}

In Fig. \ref{clCMBsyn} we report the power spectra of the $S_{G}$ and 
$S_{B}$ models compared to the CMB; in both panels, showing $E$ 
and $B$ modes, the dotted curve represents the CMB, while the 
upper and lower solid lines represent the $S_{G}$ and $S_{B}$ 
synchrotron models, respectively. 
The plots are at 100 GHz, in antenna temperature; at lower frequencies, 
the steep synchrotron frequency scaling makes the contamination rapidly 
worse, by a factor of about $[100/\nu\ (GHz)]^{5.5}$, according to the 
scaling represented in figure \ref{synspecind}. At higher frequencies, 
the contribution from polarised dust emission is likely to become 
relevant (Prunet \& Lazarian 2002). As a general feature, it can be 
noted how the Galactic emission has almost equal power on $E$ and 
$B$ modes, according to the most natural expectation for a 
non-cosmological signal (Zaldarriaga 2001). 

The contamination is severe on large angular scales, almost covering the 
reinsertion bump, in both cases. The $S_{B}$ signal drops rapidly below 
the CMB at $\ell\simeq 40$; therefore, according to this scenario, 
the synchrotron contamination to the CMB $E$ mode acoustic oscillations 
should be irrelevant. On the other hand, the $S_{G}$ model predicts 
a severe contamination also for the first CMB $E$ acoustic oscillation, 
being although irrelevant at higher multipoles. This behavior 
clarifies the qualitative difference between the two template images 
in figure \ref{synsky}, where the $S_{B}$ power has clearly the dominant 
power on large angular scales, while the fine structure contribution 
for the $S_{G}$ case is much more important. 

CMB $B$ modes are clearly dominated by foreground emission, on all 
scales and according to both models. Note that the level of $B$ power 
that we adopt is optimistically close to the upper limit set by WMAP, 
and corresponds to high energy inflation models. As we already stressed, 
at higher frequencies the contribution from polarised dust emission 
is likely to become relevant (Lazarian \& Prunet 2002), and 
even a percent polarisation degree would greatly dominate the 
CMB $B$ modes. 

Future CMB polarisation measurements must take into account the 
expected, severe contamination from diffuse Galactic synchrotron. 
In particular, $B$ modes are likely to receive a major contamination 
on all scales and at all frequencies. A way out is certainly the 
scanning of particularly clean regions; alternatively, the data 
analysis can suitably exploit the multifrequency coverage to 
isolate and subtract the Galactic contamination to the CMB. 
Although these are non-trivial tasks, they have to be accomplished 
in order to extract the whole scientific information contained in 
CMB polarisation. 

The author warmly thanks Giovanna Giardino for useful discussions. 
The HEALPix pixelisation scheme, available at {\tt www.eso.org/healpix}, 
by A.J. Banday, M. Bartelmann, K.M. Gorski, F.K. Hansen, E.F. Hivon, 
and B.D. Wandelt, has been extensively used. 

\end{document}